\documentclass[conference]{IEEEtran}
\IEEEoverridecommandlockouts
\usepackage{cite}
\usepackage{amsmath, amssymb, amsfonts}
\usepackage{algorithmic}
\usepackage{graphicx}
\usepackage{textcomp}
\usepackage{xcolor}
\usepackage{graphicx}
\usepackage{epstopdf}
\usepackage{algorithm}
\usepackage{algorithmic}
\usepackage{bbm}
\usepackage{bm}
\usepackage{mathrsfs}
\usepackage{graphicx}
\usepackage{color}
\usepackage{subcaption}
\def\BibTeX{{\rm B\kern-.05em{\sc i\kern-.025em b}\kern-.08em
 T\kern-.1667em\lower.7ex\hbox{E}\kern-.125emX}}
\begin{document}

\title{UAV Trajectory Optimization for Sensing Exploiting Target Location Distribution Map}

\author{\IEEEauthorblockN{Xiangming Du, Shuowen Zhang, and Liang Liu}
\IEEEauthorblockA{Department of Electrical and Electronic Engineering, The Hong Kong Polytechnic University\\
E-mail: xiangming.du@connect.polyu.hk, shuowen.zhang@polyu.edu.hk, liang-eie.liu@polyu.edu.hk}\thanks{This work was supported in part by the National Natural Science Foundation of China under Grant 62101474, and in part by the General Research Fund from the Hong Kong Research Grants Council under Grant 15230022.}

\vspace{-8mm}

}
\maketitle

\begin{abstract}
In this paper, we study the trajectory optimization of a cellular-connected unmanned aerial vehicle (UAV) which aims to sense the location of a target while maintaining satisfactory communication quality with the ground base stations (GBSs). In contrast to most existing works which assumed the target's location is known, we focus on a more challenging scenario where the exact location of the target to be sensed is \emph{unknown} and \emph{random}, while its distribution is known \emph{a priori} and stored in a novel \emph{target location distribution map}. Based on this map, the probability for the UAV to successfully sense the target can be expressed as a function of the UAV's trajectory. We aim to optimize the UAV's trajectory between two pre-determined locations to maximize the \emph{overall sensing probability} during its flight, subject to a GBS-UAV communication quality constraint at each time instant and a maximum mission completion time constraint. Despite the non-convexity and NP-hardness of this problem, we devise three high-quality suboptimal solutions tailored for it with polynomial complexity. Numerical results show that our proposed designs outperform various benchmark schemes.
\end{abstract}

\section{Introduction}
Unmanned aerial vehicles (UAVs) have been widely used in myriad applications due to their high mobility. To ensure the safety of UAVs, an efficient solution is \emph{cellular-enabled UAV communication} or \emph{cellular-connected UAV}, where UAVs act as a new type of aerial users served by the ground base stations (GBSs) in the cellular network \cite{ShuowenCellular}. Compared with traditional Wi-Fi based UAV communication, cellular-enabled UAV communication can extend the service range from visual line-of-sight (VLoS) to beyond VLoS, thus supporting much longer flying distance and much wider application scenarios.

Motivated by the emergence of new applications which require the \emph{sensing} function, the role of UAV in performing sensing tasks has recently attracted significant research attention \cite{Mengpaper,GuoUAV, xujiepaper,Hupaper}. Specifically, by exploiting the UAV's flexibility in the three-dimensional (3D) space, enhanced sensing performance can be achieved via proper design of the UAV's trajectory. However, existing studies typically considered the ideal scenario where the exact locations of the targets to be sensed are \emph{known}. For instance, \cite{GuoUAV} considered the case where multiple known target locations need to be visited for sensing; \cite{xujiepaper} aimed to ensure that sufficient power is radiated to every target location. However, in practice, such exact location information may not be available. Moreover, the target may appear at different locations with \emph{distinct probabilities}, which cannot be characterized by existing models and studies.

In this paper, we consider a practical and challenging scenario where the exact target location is \emph{unknown} and \emph{random}, while its statistical distribution is known \emph{a priori} based on empirical measurements or target movement pattern \cite{Chanpaper,Kaiyuepaper,Chan2,Kaiyue2}. Specifically, we propose a novel \emph{target location distribution map} to characterize and store the probabilities of appearance of the target over a geographical region, which can also quantify the probabilities for the UAV to successfully sense the target when it flies to the waypoint near each possible location. Based on this, how to design the UAV's trajectory for maximizing the overall probability of successfully sensing a target during its flight is a new challenging problem, which requires careful exploitation of the target location distribution map such that the UAV can prioritize its flight near highly-probable target locations. Moreover, how to guarantee satisfactory communication quality with the GBSs while performing the UAV's sensing task efficiently is also an open problem, which requires the trajectory design to strike the optimal balance among communication, sensing, and mission completion performances.

To address the above problems, we study the trajectory optimization of a cellular-connected UAV which bears a mission of sensing the location of a target based on its prior distribution information stored in the \emph{target location distribution map}, while maintaining satisfactory communication quality with the GBSs for ensuring its safety. We consider a general channel with potential obstructions between the UAV and GBSs, and propose to adopt the \emph{radio map} technique for characterizing the expected signal-to-noise ratio (SNR) at each UAV's possible location \cite{Radio}. We aim to optimize the UAV's trajectory to maximize the \emph{total (overall) probability} for successfully sensing the target, subject to a minimum expected SNR threshold at each time instant and a mission completion time constraint. This problem is non-convex and NP-hard. By exploiting the unique structures of the problem, we propose three algorithms for finding high-quality suboptimal solutions with polynomial complexity. Numerical results show that our proposed designs achieve significantly increased total sensing probability compared to various benchmark designs.

\section{System Model}
We consider a cellular-connected UAV which bears a mission of sensing a target while flying from an initial location $U_{\mathrm{S}}$ to a final location $U_{\mathrm{F}}$. Under a 3D Cartesian coordinate system, let $(x_{\mathrm{S}}, y_{\mathrm{S}}, H)$ and $(x_{\mathrm{F}}, y_{\mathrm{F}}, H)$ in meters (m) denote the location coordinates of $U_{\mathrm{S}}$ and $U_{\mathrm{F}}$, respectively. To ensure the safety of the UAV, the UAV needs to maintain satisfactory connectivity with the ground via communicating with a GBS in the cellular network. Let $M\geq 1$ denote the number of GBSs that are available for communication. For the purpose of drawing essential insights, we assume that the UAV flies at a constant altitude of $H$ m with constant speed $V$ meters/second (m/s), and let $\mathcal{U}\subset \mathbb{R}^{2\times 1}$ denote the feasible region of the UAV's flight projected to the horizontal plane. Let $\bm{u}(t)=[x(t),y(t)]^T \in \mathcal{U}$ denote the UAV's horizontal location at each time instant $t$, and $T$ denote the UAV's mission completion time. We aim to optimize the UAV's horizontal trajectory $\{\bm{u}(t), 0\leq t\leq T\}$ to maximize the sensing performance, subject to a communication quality constraint at each time instant during the flight, and a maximum mission completion time threshold denoted by $\bar T$ s which is equivalent to a maximum flying distance threshold given by $\bar D= V\bar T$ m.
\vspace{-6mm}
\subsection{GBS-UAV Communication Model}
\vspace{-1mm}
At every time instant during the UAV's flight, the UAV needs to conduct control and non-payload communication with one of the GBSs for ensuring its safety\cite{ShuowenCellular}, which requires low data volume and high reliability, thus single-stream transmission is preferred regardless of the numbers of antennas at the GBS and the UAV. In this paper, we focus on downlink communication from the GBS to the UAV, while our results are also directly applicable to the case of uplink communication. Denote the effective channel gain from the $m$-th GBS to the UAV at horizontal location $\bm{u}$ as $g_m(\bm{u})=\bar g_m(\bm{u})\tilde g_m(\bm{u})\in \mathbb{R}$, where $\bar g_m(\bm{u})\in \mathbb{R}$ and $\tilde g_m(\bm{u})\in \mathbb{R}$ denote the large-scale channel gain and the small-scale fading gain, respectively, with $\mathbb{E}[\tilde {g}_m^2(\bm{u})]=1$. Specifically, $\bar g_m(\bm{u})$ consists of the path loss, shadowing, and antenna gains at the GBS and the UAV, thus being a static function of the location of the $m$-th GBS, $\bm{u}$, and the UAV's altitude $H$. Therefore, $\bar g_m(\bm{u})$ for any $\bm{u}\in \mathcal{U}$ can be measured or calculated prior to the UAV's flight \cite{Radio}. On the other hand, $\tilde g_m(\bm{u})$ is determined by the real-time small-scale fading which changes rapidly over channel coherence intervals. Denote $P_m$ as the transmit power at the $m$-th GBS, and $\sigma^2$ as the effective noise power at the UAV receiver. The received SNR at the UAV if the $m$-th GBS is selected for transmission is given by $\rho_m(\bm{u})=\frac{P_mg^2_m(\bm{u})}{\sigma^2}=\frac{P_m\bar g^2_m(\bm{u})\tilde g^2_m(\bm{u})}{\sigma^2}$.

Since the small-scale fading gain $\tilde g_m(\bm{u})$ is generally a random variable and cannot be known prior to the UAV's flight, we adopt the \emph{expected SNR} as the communication performance metric in the trajectory optimization, and consider a minimum threshold for it denoted by $\bar \rho$.\footnote{Note that in $\rho_m(\bm{u})$, we did not consider real-time beamforming at the GBS/UAV based on the instantaneous small-scale channel. If such beamforming is considered, $\rho_m(\bm{u})$ and the expected SNR can be further improved; the expected SNR is still guaranteed to be higher than the required threshold $\bar \rho$.} Under this metric, the GBS with the highest expected SNR should be associated with the UAV for communication, which leads to a resulting expected SNR at the UAV given by $\bar \rho(\bm{u})=\max\limits_{m\in \mathcal{M}}\mathbb{E}[\rho_m(\bm{u})]=\max\limits_{m\in \mathcal{M}}\frac{P_m\bar {g}_m^2(\bm{u})}{\sigma^2}\geq \bar\rho, \forall \bm{u}=\bm{u}(t),0\leq t\leq T$.

Note that the expected SNR $\bar\rho(\bm{u})$ is determined by $\bar g_m(\bm{u})$, which consists of the shadowing effect and is critically dependent on the terrain features (e.g., location, height, and shape of the obstacles). In general, it is difficult to analytically model $\bar g_m(\bm{u})$ and consequently $\rho (\bm{u})$ as explicit and tractable functions of $\bm{u}$ to facilitate trajectory optimization. In this paper, we adopt a \emph{map-based approach} to characterize the expected SNR, where the values of $\bar\rho(\bm{u})$ for all $\bm{u}\in \mathcal{U}$'s are stored in an \emph{expected SNR map}\cite{Radio}. Specifically, we first quantize the continuous region $\mathcal{U}$ into $D \times D$ square grids each with length $\Delta_{\mathrm{D}}$. For simplicity, we assume $\mathcal{U}$ is an $L$ m $\times L$ m square region and $D=\frac{L}{\Delta_{\mathrm{D}}}$. The quantization granularity $\Delta_{\mathrm{D}}$ is selected as a sufficiently small value such that the large-scale channel gain and consequently the expected SNR remains approximately constant in each grid. Thus, all locations in each $(i,j)$-th grid can be well-represented by the grid center (or ``grid point'') denoted by $\bm{u}_{\mathrm{D}}(i, j)=[i - \frac{1}{2}, j - \frac{1}{2}]^T\Delta_{\mathrm{D}}, i, j\in \mathcal{D}$ with $\mathcal{D}=\{1, ..., D\}$. Based on this, we can use a $D\times D$ matrix to store the expected SNR values at all grid points denoted by $\bm{S}\in \mathbb{R}^{D\times D}$, where each $(i,j)$-th element is given by $
[\bm{S}]_{i,j}=\bar \rho(\bm{u}_{\mathrm{D}}(i, j)) = \max\limits_{m\in \mathcal{M}}\frac{P_m\bar {g}_m^2(\bm{u}_{\mathrm{D}}(i, j))}{\sigma^2},i,j\in \mathcal{D}$. It is worth noting that $\bm{S}$ can be efficiently obtained via measurement or ray-tracing methods prior to the UAV's flight\cite{Radio}. By depicting $\bm{S}$ for all grid points (i.e., all $i,j\in \mathcal{D}$), we have a so-called \emph{expected SNR map}. In Fig. 1(a), we illustrate an expected SNR map under the setup in Section V.

Based on the above and by noting that $\Delta_{\mathrm{D}}$ is sufficiently small, we propose a \emph{discretized trajectory structure} for the UAV. Specifically, the path of the UAV is composed of connected line segments, where the two end points of each line segment are two adjacent grid points, i.e., those with a distance no larger than $\sqrt{2}\Delta_{\mathrm{D}}$. Therefore, the UAV's trajectory can be represented by a series of grid points $\{\bm{u}_{\mathrm{D}}(i_n, j_n)\}_{n=1}^{N}$, where $i_n, j_n\in \mathcal{D}$ and $N$ denotes the total number of points. Consequently, to satisfy the expected SNR constraint, $[\bm{S}]_{i_n,j_n}\geq \bar\rho, \forall n $ should hold. Note that as $\Delta_{\mathrm{D}} \rightarrow 0$, the discretized trajectory approaches the continuous trajectory.
\vspace{-3mm}
\subsection{Target Sensing Model}
\vspace{-1mm}
In the target sensing mission, the UAV aims to sense the location of a target on the ground. Specifically, the horizontal location of the target denoted by $\bm{u}_{\mathrm{T}}=[x_{\mathrm{T}},y_{\mathrm{T}}]^T$ is \emph{unknown} and \emph{random}, while its spatial distribution over the two-dimensional (2D) space is known \emph{a priori} for exploitation, which can be obtained based on empirical data or target movement pattern. Let $p_{x_{\mathrm{T}}, y_{\mathrm{T}}}(x_{t}, y_{t})$ denote the probability density function (PDF) for the target's horizontal location $\bm{u}_{\mathrm{T}}$, where $[x_{t}, y_{t}]^T\in \mathcal{U}$. We assume that the UAV is able to sense the target if the target's horizontal location lies in the same grid as the UAV, i.e., $|\bm{u}_{\mathrm{D}}(i_n, j_n)-\bm{u}_{\mathrm{T}}|\preceq [\frac{\Delta_{\mathrm{D}}}{2}, \frac{\Delta_{\mathrm{D}}}{2}]^T$.\footnote{Note that this model is applicable to various sensing methods. For camera-based sensing where the UAV senses a target by capturing video/image of it, this can guarantee a sufficiently high resolution. For radar sensing where the UAV is equipped with multiple antennas that constitute multiple-input multiple-output (MIMO) radar over non-overlapping frequency bands with GBS-UAV communication, this can guarantee that the received echo signal has sufficiently strong power, which leads to a sufficiently low sensing mean-squared error (MSE). It is also worth noting that the sensing accuracy can always be improved with a smaller grid granularity, such that the UAV is closer to the continuous locations in a grid, at a cost of higher complexity in map storage and trajectory optimization.} In this case, the probability for the UAV to sense a target when it is located at $\bm{u}_{\mathrm{D}}(i_n, j_n)$ is given by the integral of the probabilities over all possible target locations in the corresponding grid, namely, $\int_{(i_n-1)\Delta_{\mathrm{D}}}^{i_n\Delta_{\mathrm{D}}}\int_{(j_n-1)\Delta_{\mathrm{D}}}^{j_n\Delta_{\mathrm{D}}}p_{x_{\mathrm{T}}, y_{\mathrm{T}}}(x_{t}, y_{t})dy_{t}dx_{t}$.

We aim to design the UAV's trajectory to maximize the \emph{total probability} of sensing the target during the flight, which is given by $\sum_{n=1}^N\int_{(i_n-1)\Delta_{\mathrm{D}}}^{i_n\Delta_{\mathrm{D}}}\int_{(j_n-1)\Delta_{\mathrm{D}}}^{j_n\Delta_{\mathrm{D}}}p_{x_{\mathrm{T}}, y_{\mathrm{T}}}(x_{t}, y_{t})dy_{t}dx_{t}$. To this end, we introduce a \emph{target location distribution map}, which is represented by a matrix denoted by $\bm{P}\in \mathbb{R}^{D\times D}$ consisting of the target appearance probabilities and equivalently sensing probabilities in all $D\!\times\! D$ grids. Specifically, each $(i,j)$-th element in $\bm{P}$ is given by $[\bm{P}]_{i, j}=\int_{(i-1)\Delta_{\mathrm{D}}}^{i\Delta_{\mathrm{D}}}\int_{(j-1)\Delta_{\mathrm{D}}}^{j\Delta_{\mathrm{D}}}p_{x_{\mathrm{T}}, y_{\mathrm{T}}}(x_{t}, y_{t})dy_{t}dx_{t}, i, j\in \mathcal{D}$. Hence, the total sensing probability during the UAV's flight is $\sum_{n=1}^N[\bm{P}]_{i_n, j_n}$. By further noting that visiting a grid point more than once is not beneficial to sensing, communication, or mission completion performance, we consider a non-repeated flight where each grid point is visited at most once, i.e., $\bm{u}_{\mathrm{D}}(i_n, j_n) \ne \bm{u}_{\mathrm{D}}(i_m, j_m), \forall n\ne m, n, m\in \mathcal{N}\overset{\Delta}{=} \{1, ..., N\}$.

\emph{Remark (Example of a Target Location Distribution):} A practical target location distribution model is the 2D Gaussian mixture model, where the PDF is the weighted sum of $S\!\geq\! 1$ 2D Gaussian PDFs, each with mean $(x_s,y_s)$, variance $\sigma_{s}^2$, and weight $p_s\!\in\! [0,1]$ that satisfies $\sum_{s=1}^Sp_s\!=\!1$, namely, $p_{x_{\mathrm{T}}, y_{\mathrm{T}}}(x_{t}, y_{t})\!=\!\sum_{s=1}^S\frac{p_s}{2\pi\sigma_{s}^2}e^{-\frac{(x_{t}-x_{s})^2+(y_{t}-y_{s})^2}{2\sigma_{s}^2}}$. In Fig. 1(b), we illustrate a target location distribution map $\bm{P}$ under this model, with $\Delta_{\mathrm{D}}\!=\!30$ m, $p_1\!=\!p_2\!=\!0.5$, $\sigma_1\!=\!1.8\Delta_{\mathrm{D}}$, $\sigma_2\!=\!2\Delta_{\mathrm{D}}$, $(x_1,y_1)\!=\!(13\Delta_{\mathrm{D}}, 5\Delta_{\mathrm{D}})$, and $(x_2,y_2)\!=\!(6\Delta_{\mathrm{D}}, 15\Delta_{\mathrm{D}})$.
\vspace{-1mm}
\section{Problem Formulation}
\vspace{-1mm}
We aim to optimize the UAV's trajectory to maximize the total sensing probability, subject to an expected SNR constraint at each time instant during the flight and a maximum flying distance constraint. For ease of exposition, we assume that $\bm{u}_{\mathrm{S}}\!\overset{\Delta}{=}\![x_{\mathrm{S}},y_{\mathrm{S}}]^T$ and $\bm{u}_{\mathrm{F}}\!\overset{\Delta}{=}\![x_{\mathrm{F}},y_{\mathrm{F}}]^T$ are grid points.\footnote{If $\bm{u}_{\mathrm{S}}$ and $\bm{u}_{\mathrm{F}}$ are not grid points, we can let the UAV firstly fly to the nearest grid point from $\bm{u}_{\mathrm{S}}$ and lastly fly from the nearest grid point to $\bm{u}_{\mathrm{F}}$.} Under the discretized trajectory structure, the problem is formulated as\vspace{-3pt}
\setlength{\jot}{0pt}
\begin{align} \mbox{(P1)}\!\!\!
\mathop{\max_{\{i_n, j_n\}_{n=1}^N}} \ &\sum\limits_{n=1}^{N} [\bm{P}]_{i_n, j_n}\\[-0.5mm]
\mathrm{s.t.} \quad &\bm{u}_{\mathrm{D}}(i_1, j_1) =\bm{u}_{\mathrm{S}}, \bm{u}_{\mathrm{D}}( i_N, j_N) =\bm{u}_{\mathrm{F}}\label{P1_C1}\\[-0.5mm]
&[\bm{S}]_{i_n,j_n}\geq\bar \rho, \forall n\in\mathcal{N} \label{P1_C2}\\[-0.5mm]
&\sum\limits_{n=1}^{N-1} | \bm{u}_{\mathrm{D}}(i_{n+1}, j_{n+1})\! - \! \bm{u}_{\mathrm{D}}(i_n, j_n)\|\! \leq \bar D\label{P1_C3}\\[-0.5mm]
&\left\| \bm{u}_{\mathrm{D}}(i_{n+1}, j_{n+1})\!-\! \bm{u}_{\mathrm{D}}(i_n, j_n)\right\|\! \leq \!\sqrt{2}\Delta_{\mathrm{D}}, \nonumber\\[-0.5mm]
&\qquad\qquad\qquad\qquad\qquad\forall n\in\mathcal{N}\label{P1_C4}\\[-0.5mm]
&\bm{u}_{\mathrm{D}}(i_n,\!j_n)\! \ne \! \bm{u}_{\mathrm{D}}(i_m,\!j_m ), \!\forall n \!\ne \! m,\!n, \!m\!\in\! \mathcal{N}\!\label{P1_C5}\\[-0.5mm]
&i_n, j_n\in \mathcal{D},\qquad\qquad\quad \ \forall n\in\mathcal{N}\label{P1_C6}.
\end{align}
Note that (P1) is a non-convex combinatorial optimization problem due to the integer optimization variables in $\{i_n, j_n\}_{n=1}^N$. Moreover, (P1) is a constrained longest path problem, which can be shown to be NP-hard\cite{KSP}. Particularly, note that to achieve a high total sensing probability, the optimal trajectory needs to traverse all grid points, which may lead to unaffordable flying distance. On the other hand, a trajectory that solely aims to traverse grid points with high sensing probabilities may be infeasible due to the expected SNR constraint. To summarize, the optimal trajectory needs to strike the best balance among the total probability, expected SNR, and flying distance, which makes (P1) very difficult to solve.
\begin{figure}[t]
	\centering
	\begin{subfigure}[b]{0.45\linewidth}
		\includegraphics[width=\linewidth]{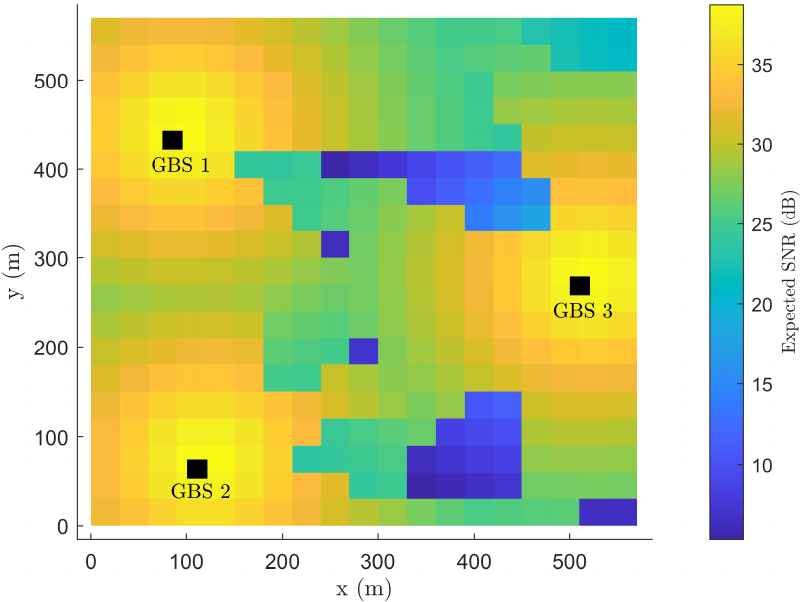}\vspace{-3pt}
		\caption{Illustration of $\bm{S}$.}
		\label{fig:sub1}
	\end{subfigure}
	\hfill
	\begin{subfigure}[b]{0.45\linewidth}
		\includegraphics[width=\linewidth]{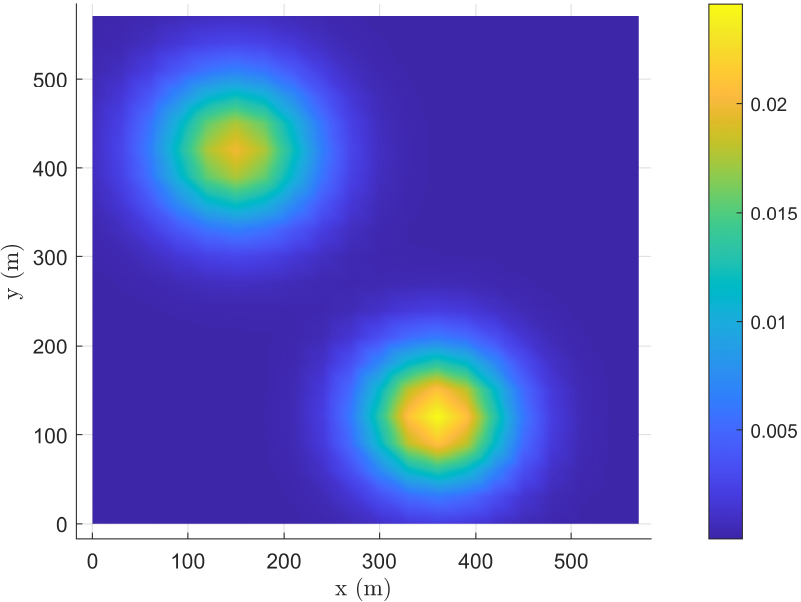}\vspace{-3pt}
		\caption{Illustration of $\bm{P}$.}
		\label{fig:sub2}
	\end{subfigure}\vspace{-6pt}
	\caption{Illustration of an expected SNR map $\bm{S}$ and a target location distribution map $\bm{P}$.}
	\label{fig:combined}\vspace{-9mm}
\end{figure}
\vspace{-1mm}
\section{Proposed Solutions}
\vspace{-1mm}
\belowdisplayskip=2pt
\subsection{Graph-based Problem Reformulation}
\vspace{-1mm}
To overcome the above challenges in solving (P1), we first propose to replace the objective function of (P1) with a lower bound of it given by\vspace{-6pt}
\begin{align}
\sum\limits_{n=1}^{N}[\bm{P}]_{i_{n}, j_{n}}\geq 1/\left( \sum\limits_{n=1}^{N}1/[\bm{P}]_{i_{n}, j_{n}}\right),\label{re-P}
\end{align}
where the inequality holds due to the relationship between geometric mean and arithmetic mean.\footnote{If $[\bm{P}]_{i,j}=0$, we can assign a small value to $[\bm{P}]_{i,j}$ to make it invertible.} Based on this, we transform (P1) into (P2) below:\vspace{-6pt}
\begin{align} \mbox{(P2)}
\mathop{\min_{\{i_n, j_n\}_{n=1}^N:(\ref{P1_C1})-(\ref{P1_C6})}} \ &\sum\limits_{n=1}^{N} 1/[\bm{P}]_{i_{n}, j_{n}}.\label{P2}
\end{align}

Next, we propose a graph-based model for (P2). Specifically, we construct an undirected weighted graph $G=(V,E)$ with two sets of weights. The vertex set $V$ is given by\vspace{-6pt}
\begin{align}
V=\{U_{\mathrm{D}}(i, j):[\bm{S}]_{i, j}\geq \bar{\rho}, i, j\in \mathcal{D}\}, \label{V}
\end{align}
where $U_{\mathrm{D}}(i, j)$ denotes the $(i, j)$-th grid point at location $\bm{u}_{\mathrm{D}}(i, j)$ that satisfies the expected SNR constraint. The edge set $E$ is given below for $(i,j)\neq(k,l)$:\vspace{-8pt}
\begin{align}
\!\!\!\!\!E\!=\! \{ \left(U_{\mathrm{D}}(i, j),\!U_{\mathrm{D}}(k, l)\right)\!\!:\!\!\left\| \bm{u}_{\mathrm{D}}(i, j)\!-\!\bm{u}_{\mathrm{D}}(k , l)\right\|\! \leq\! \sqrt{2}\Delta_{\mathrm{D}}\},\!\!\!\label{E}
\end{align}
where an edge exists between two vertices if and only if they are adjacent. For each edge, we have a \emph{distance weight} given below which represents the distance between two locations:\vspace{-8pt}
\begin{align}
W^{\mathrm{D}}\left( U_{\mathrm{D}}(i, j), U_{\mathrm{D}}(k, l\right))= \left\| \bm{u}_{\mathrm{D}}(i, j)- \bm{u}_{\mathrm{D}}(k, l)\right\|\!.\label{WD}
\end{align}
We also have a \emph{probability weight} which denotes the inverse of sensing probability at the location of the latter vertex:\vspace{-8pt}
\begin{align}
W^{\mathrm{P}}\left( U_{\mathrm{D}}(i, j), U_{\mathrm{D}}(k, l\right))= 1/{[\bm{P}]_{k, l}}.\label{WP}
\end{align}
Note that any path in $G$ from $U_{\mathrm{D}}(i_1, j_1)$ to $U_{\mathrm{D}}(i_N, j_N)$ can be characterized by a $2 \times N$ matrix $\bm{I}=[[i_1,j_1]^T, [i_2,j_2]^T, ..., [i_N,j_N]^T]$, for which the corresponding trajectory always satisfies the expected SNR constraint. The sum flying distance of $\bm{I}$ is given by\vspace{-8pt}
\begin{align}
f^{\mathrm{D}}(\bm{I})=\sum\limits_{n=1}^{N-1}{W^{\mathrm{D}}( U_{\mathrm{D}}(i_n, j_n), U_{\mathrm{D}}(i_{n+1}, j_{n+1}))}.\label{fD}
\end{align}
The sum sensing probability inverse of $\bm{I}$ is given by\vspace{-8pt}
\begin{align}
\!\!\!\!\!f^{\mathrm{P}}(\bm{I})\!=\!\!\frac{1}{[\bm{P}]_{i_{1}, j_{1}}}\!+\!\sum\limits_{n=1}^{N-1}{ W^{\mathrm{P}} ( U_{\mathrm{D}}(i_n, j_n), U_{\mathrm{D}}(i_{n+1} , j_{n+1}))}.\!\!\!\label{fP}
\end{align}
Therefore, (P2) is equivalent to the following problem:\vspace{-8pt}
\begin{align} \mbox{(P3)}
\mathop{\min_{\bm{I}:f^{\mathrm{D}}(\bm{I})\leq \bar D}} \ &f^{\mathrm{P}}(\bm{I}).\label{P3}
\end{align}

The feasibility for (P3) and consequently (P2) and (P1) can be checked by finding the shortest path from $U_{\mathrm{D}}(i_1, j_1)$ to $U_{\mathrm{D}}(i_N, j_N)$ with respect to the distance weight $f^{\mathrm{D}}(\bm{I})$ via the Dijkstra algorithm\cite{Graph}. If the resulting minimum distance is no larger than $\bar D$, (P3) and (P1) are feasible. In the following, we study (P3) assuming it has been verified to be feasible. 

Note that (P3) is still a non-convex problem due to the integer variables in $\bm{I}$. Moreover, it is a \emph{constrained shortest path problem} which is also NP-hard\cite{outage}. Finding the optimal solution to (P3) via exhaustive search requires complexity $\mathcal{O}(D^2!)$, which is unaffordable even for moderate map size. In the following, we propose a low-complexity high-quality suboptimal solution to (P3) via graph theory and convex optimization, which is also a suboptimal solution to (P1).
\vspace{-2mm}
\subsection{Proposed Solution I}
\vspace{-1mm}
In this subsection, we employ the Lagrangian relaxation method to obtain a suboptimal solution to (P3). Specifically, the Lagrangian of (P3) is given by $\mathscr{L}(\bm{I},\lambda)=f^{\mathrm{P}}(\bm{I})+\lambda f^{\mathrm{D}}(\bm{I}), $ where $\lambda$ is the dual variable. The Lagrange dual function is then given by $
g(\lambda)=\min\limits_{\bm{I}}{\mathscr{L}(\bm{I}, \lambda)}=\min\limits_{\bm{I}} f^{\mathrm{P}}(\bm{I})+\lambda f^{\mathrm{D}}(\bm{I})$. Consequently, the dual problem is given by\vspace{-5pt}
\begin{align} \mbox{(P3-Dual)}\quad
\mathop{\max_{\lambda\geq0 } \quad \min_{\bm{I}}} \quad &f^{\mathrm{P}}(\bm{I})+\!\lambda f^{\mathrm{D}}(\bm{I}).\label{P3-Dual}
\end{align}

Note that the dual problem (P3-Dual) is a convex optimization problem. However, the duality gap between (P3) and (P3-Dual) is generally non-zero, due to the non-convexity of (P3). Inspired by \cite{outage} which deals with a similar problem, we propose to solve (P3-Dual) and further find a high-quality primal solution denoted by $\bm{I}_{\mathrm{I}}$ via subgradient-based method and $K$-shortest path algorithm\cite{outage}. The details of this algorithm can be found in \cite{outage}. Note that this algorithm is guaranteed to obtain a feasible solution to (P3) and (P1). The worst-case complexity of the algorithm can be shown to be $\mathcal{O}(D^8\log^2D^2+D^6K)$, which is significantly reduced compared to that for finding the optimal solution, $\mathcal{O}(D^2!)$\cite{outage}. 

Note that due to the replacement of objective function in (P3), the optimal solutions to (P1) and (P3) may not be the same, which may also lead to a difference between proposed solution I to (P3) and the optimal solution to (P1). In the following, we aim to mitigate such difference by proposing two further-improved solutions tailored to the structure of (P1).
\vspace{-7mm}
\subsection{Proposed Solution II}
\vspace{-1mm}
Although proposed solution I provides a systematic approach of finding a feasible solution to (P1), the minimizing nature of the transformed problem (P3) tends to reduce the number of grids visited by the UAV, which may limit the total sensing probability. For example, some grids with high sensing probabilities may be missed in proposed solution I. 

To address this issue, we propose to utilize proposed solution I as an \emph{initial trajectory}, and improve it by allowing the UAV to deviate from it at one waypoint and fly to a nearby grid point with high sensing probability before completing the flight. \emph{Firstly}, we identify all the feasible grid points which satisfy the expected SNR constraint but were not selected in the initial trajectory and let $V_{\mathrm{F}}$ denote its set. \emph{Secondly}, we sort them in a decreasing order of their corresponding sensing probabilities, and select the top $R_{\mathrm{I}}\geq 1$ ones with the highest probabilities, which are denoted by $\{U_{\mathrm{D}}(k_r, l_r)\}_{r=1}^{R_{\mathrm{I}}}$ in graph $G$. \emph{Thirdly}, for each $U_{\mathrm{D}}(k_r, l_r)$, we find its nearest location in the initial trajectory denoted by $U_{\mathrm{D}}(k'_r, l'_r)$ in $G$. Based on this, we propose a new trajectory for each $r$ denoted by $\bm{I}_r$ that consists of three parts: 1) $U_{\mathrm{D}}(i_1,j_1)$ to $U_{\mathrm{D}}(k'_r,l'_r)$ same as the initial trajectory; 2) $U_{\mathrm{D}}\hspace{-0.5pt}(k'_r,l'_r)$ to $U_{\mathrm{D}}(k_r,l_r)$ as the shortest-distance trajectory under expected SNR constraint obtained via the Dijkstra algorithm over graph $G$; 3) $U_{\mathrm{D}}(k_r, l_r)$ to $U_{\mathrm{D}}(i_N, j_N)$ as the shortest-distance trajectory under expected SNR constraint obtained via the Dijkstra algorithm over graph $G$. For each $\bm{I}_r$, we obtain the sum flying distance $f^{\mathrm{D}}(\bm{I}_r)$ according to (\ref{fD}). If we cannot find a trajectory that satisfies the above requirements, we set $f^{\mathrm{D}}(\bm{I}_r)=\infty$. \emph{Finally}, among all $\bm{I}_r$'s and the initial trajectory $\bm{I}_\mathrm{I}$, we select the best trajectory as the one with a sum flying distance no larger than $\bar D$ and a maximum total sensing probability (the original objective function of (P1)). The obtained solution denoted by $\bm{I}_\mathrm{II}$ (proposed solution II) is guaranteed to achieve no smaller total sensing probability compared to $\bm{I}_\mathrm{I}$ due to the above selection procedure. The worst-case complexity for obtaining $\bm{I}_\mathrm{II}$ based on $\bm{I}_\mathrm{I}$ can be shown to be $\mathcal{O}(D^2+R_\mathrm{I}(2D^4))$. Note that as $R_\mathrm{I}$ increases, the performance will increase at a cost of higher complexity. Thus, the value of $R_\mathrm{I}$ can be flexibly chosen according to practical requirements.
\vspace{-2mm}
\subsection{Proposed Solution III}
\vspace{-1mm}
In proposed solution III, we further enhance the performance by including \emph{multiple} extra waypoints and allowing more flexible waypoint visiting order. Consider proposed solution I as the \emph{initial trajectory}. \emph{Firstly}, we select the top $R_{\mathrm{II}}\geq 1$ feasible grid points in $V_{\mathrm{F}}$ with the highest probabilities. \emph{Secondly}, we aim to construct a new trajectory from the initial location $U_{\mathrm{D}}(i_1,\! j_1)$ to the final location $U_{\mathrm{D}}(i_N,\!j_N)$ in $G$ which traverses all waypoints in the initial trajectory and all $R_{\mathrm{II}}$ new points with highest sensing probabilities. In light of the maximum flying distance constraint, this trajectory is designed to minimize the sum flying distance, which corresponds to a traveling salesman problem (TSP). Although TSP is an NP-hard problem, various algorithms have been developed for finding a high-quality suboptimal solution, such as the ant colony optimization (ACO) algorithm. Let $\bm{I}_{R_{\mathrm{II}}}$ denote the obtained trajectory with $R_{\mathrm{II}}$ more waypoints. If no feasible $\bm{I}_{R_{\mathrm{II}}}$ exists or $f^{\mathrm{D}}(\bm{I}_{R_{\mathrm{II}}})\!\!>\!\!\bar D$, we will repeat the above procedures by incorporating one fewer, i.e., $R_{\mathrm{II}}\!-\!1$, new waypoints with highest sensing probabilities, until a feasible solution is obtained or the number of new waypoints is reduced to zero (i.e., the trajectory is the same as the initial trajectory). The resulting proposed solution III denoted by $\bm{I}_{\mathrm{III}}$ is guaranteed to achieve no smaller total sensing probability than the initial trajectory due to the above selection procedure. The worst-case complexity for proposed solution III via ACO for TSP can be shown to be $\mathcal{O}(R_{\mathrm{II}}(CA(R_{\mathrm{II}}+N)^2+(R_{\mathrm{II}}+N)D^4)+(R_{\mathrm{II}}+N)^2D^4)$, where $C$ is the number of iterations and $\!A\!$ is the number of ants in ACO.

\vspace{-2mm}
\section{Numerical Results}
\vspace{-1mm}
We consider a scenario shown in Fig. 2(a), with $L\!=\!600$ m and $M\!=\!3$ GBSs which have common height of $10$ m and same transmit power of $P_m\!=\!25$ dBm, $\forall m$. The large-scale channel is modeled under the urban micro (UMi) setup specified in 3GPP. The UAV flies at an altitude of $H = 80$ m and has an average receiver noise power of $\sigma^2\!=\!-90$ dBm. The grid granularity is set as $\Delta_{\mathrm{D}}\!=\!30$ m if not specified otherwise. The expected SNR map is shown in Fig. 1(a). For the target location distribution, we consider a truncated version of the map shown in Fig. 1(b) under Gaussian mixture PDF. Specifically, we remove the grids with obstacles and normalize the sensing probabilities of the entire area such that the sum sensing probabilities over all grids that do not overlap with obstacles is still $1$. We consider an expected SNR target of $\bar \rho\!=\!7$ dB, under which the infeasible grid points are shown in Fig. 2(a). We consider a benchmark scheme where the UAV flies in the \emph{shortest-distance trajectory} under the expected SNR constraint, without considering the sensing probability\cite{Radio}.

In Fig. 2(a), we show the trajectory designs via our proposed solutions I, II, III and the benchmark scheme under $\bar D\!=\! 2700$ m. It is observed that our proposed solutions tend to traverse grids with high sensing probabilities at the cost of higher flying distance compared to the benchmark scheme. In Fig. 2(b), we show the total sensing probability versus the maximum flying distance $\bar D$ for the aforementioned schemes, as well as proposed solutions II and III with the benchmark scheme (shortest-distance trajectory) as the initial trajectory. It is observed that all the proposed solutions outperform the benchmark scheme, and both proposed solutions II and III outperform proposed solution I due to the further improvements. Moreover, when the flying distance constraint is tight (i.e., $\bar D$ is small), proposed solution II outperforms proposed solution III since the limited flying distance may not allow the inclusion of many high-probability locations; while when the flying distance constraint becomes relaxed (i.e., $\bar D$ is large), proposed solution III outperforms proposed solution II since it incorporates more high-probability locations with flexible visiting order design. In this setup, proposed solution III can improve the overall probability from $0.1649$ (proposed solution II) to $0.7404$ with $\bar D=2700$ m. Finally, it is observed that for proposed solutions II and III, using proposed solution I as the initial trajectory generally leads to improved performance compared with using the shortest-distance trajectory, due to the joint consideration of the sensing probability and flying distance via the Lagrange relaxation method. In addition, we evaluate the performance under $\Delta_{\mathrm{D}}\!=\!60$ m in Fig. 2(c), where the performance is observed to be worse than the case with $\Delta_{\mathrm{D}}\!=\!30$ m. In Fig. 2(d), we show the computation time of different designs, where the case with $\Delta_{\mathrm{D}}\!=\!60$ m is observed to consume less computation time, thus validating the performance-complexity trade-off in selecting $\Delta_{\mathrm{D}}$.

\begin{figure}[t]
	\centering
	\begin{subfigure}[b]{1.7in}
		\includegraphics[width=1.4in]{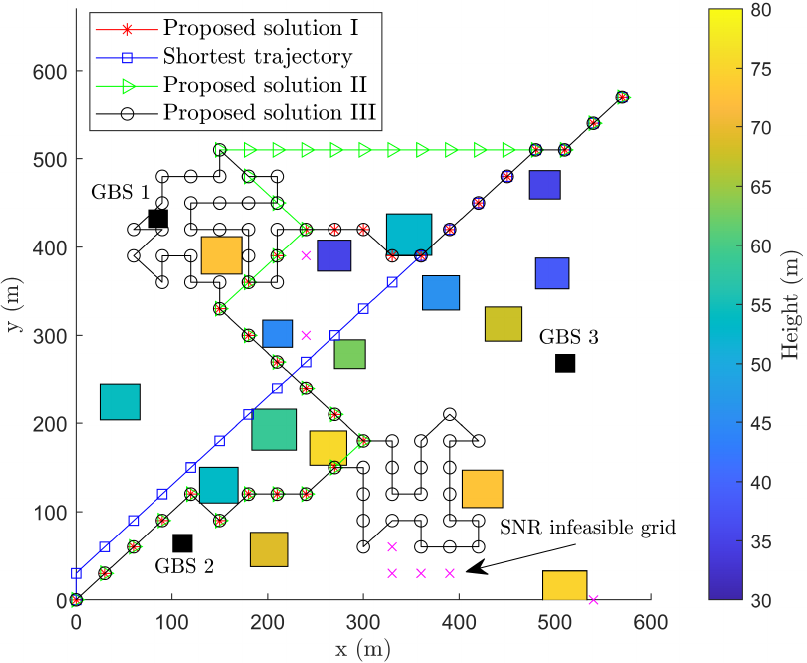}
		\vspace{-2mm}
		\caption{Trajectory designs with $\bar D= 2700$ m,  $\Delta_{\mathrm{D}}=30$ m.}
	\end{subfigure}
	\begin{subfigure}[b]{1.7in}
		\includegraphics[width=1.3in]{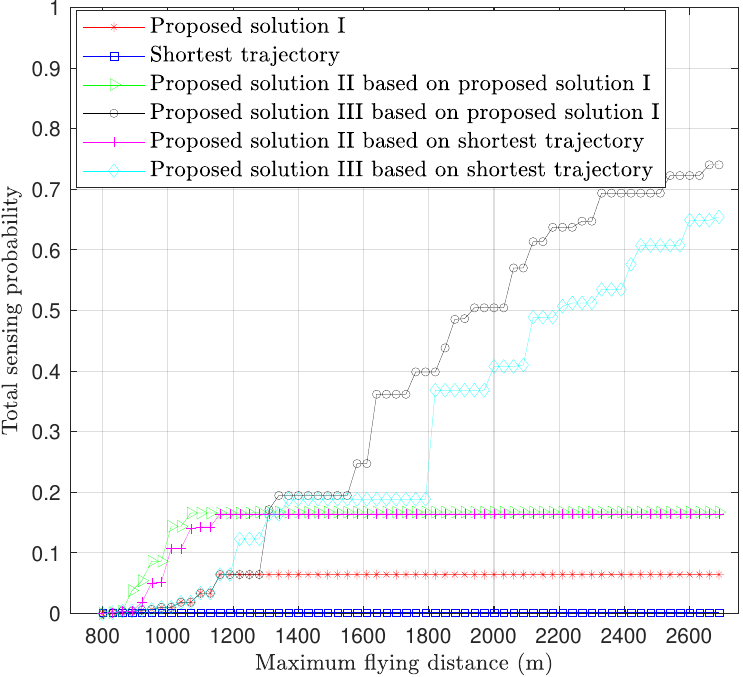}
		\vspace{-2mm}
		\caption{Total sensing probability versus $\bar D$ with $\Delta_{\mathrm{D}}=30$ m.}
	\end{subfigure}
	\begin{subfigure}[b]{1.7in}
		\includegraphics[width=1.4in]{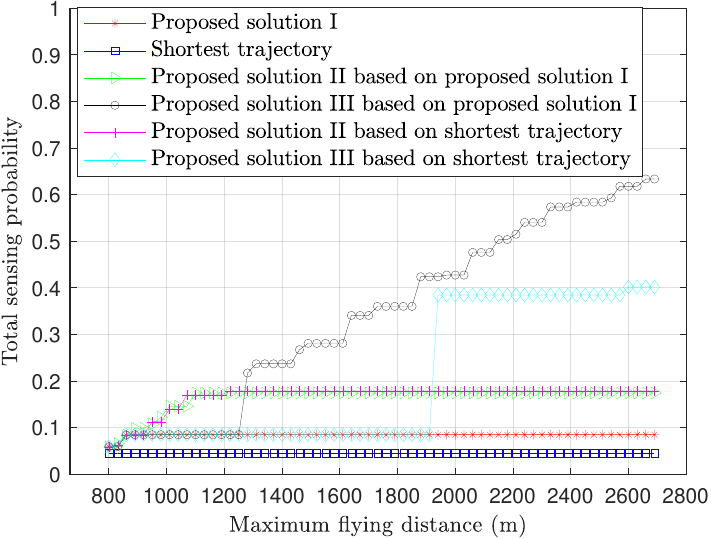}
		\vspace{-2mm}
		\caption{Total sensing probability versus $\bar D$ with $\Delta_{\mathrm{D}}=60$ m.}
	\end{subfigure}
	\begin{subfigure}[b]{1.7in}
		\includegraphics[width=1.4in]{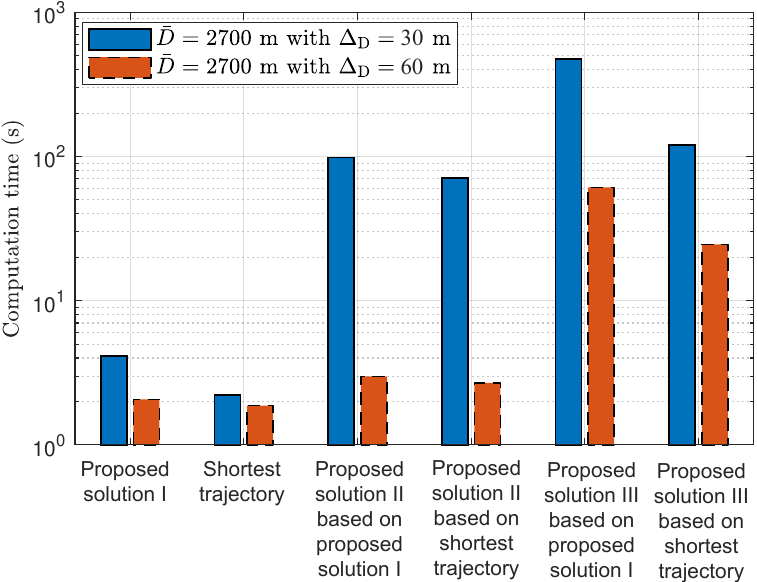}
		\vspace{-2mm}
		\caption{Computation time with $\bar D= 2700$ m.}
	\end{subfigure}
	\vspace{-2mm}
	\caption{{Illustration and performance of trajectory designs.}}
	\vspace{-8mm}
\end{figure}

\vspace{-1mm}
\section{Conclusions}
\vspace{-1mm}
This paper studied the trajectory optimization of a cellular-connected UAV in a sensing mission. Under a challenging scenario where the location of the target is unknown and random, we quantified the successful sensing probability at each possible UAV location, and studied the trajectory optimization problem to maximize the total sensing probability over the flight, subject to a communication quality constraint and a mission completion time constraint. In the future, it is worthwhile extending our studies to more general 3D trajectory optimization with advanced GBS-UAV communication techniques such as multi-GBS cooperative communication.

\bibliographystyle{IEEEtran}
\bibliography{reference}
\end{document}